\documentclass[twocolumn]{aa}
\usepackage{graphics}
\input{epsf}
\newcommand{\bfo}[1]{\mbox{\boldmath $#1$}}
\def\bvarphi{\mbox{\boldmath $\varphi$}}

\begin{document}
\newcommand{\beq}{\begin{equation}}
\newcommand{\eeq}{\end{equation}}
\def\la{\hbox{\raise.35ex\rlap{$<$}\lower.6ex\hbox{$\sim$}\ }}
\def\ga{\hbox{\raise.35ex\rlap{$>$}\lower.6ex\hbox{$\sim$}\ }}
\def\runit{\hat {\bf  r}}
\def\phunit{\hat {\bfo \bvarphi}}
\def\zunit{\hat {\bf z}}
\def\beq{\begin{equation}}
\def\eeq{\end{equation}}
\def\beqa{\begin{eqnarray}}
\def\eeqa{\end{eqnarray}}
\def\sub#1{_{_{#1}}}
\def\order#1{{\cal O}\left({#1}\right)}
\newcommand{\sfrac}[2]{\small \mbox{$\frac{#1}{#2}$}}
\title{Hydrodynamic response of rotationally supported flows
in the Small Shearing Box model
}

\author{A. Sternberg$^1$, O.M. Umurhan$^{1,2,3}$, Y. Gil$^1$ \and O. Regev$^{1,4}$}

\offprints{A. Sternberg, \email{phassaf@techunix.technion.ac.il}}

\institute{$^1$Department of Physics, Technion-Israel Institute of
Technology,
  32000 Haifa, Israel \\
  $^2$Department of Geophysics and Space Sciences, Tel-Aviv
University,  Tel-Aviv, Israel \\
  $^3$Department of Astronomy, City College of San Francisco,
  San Francisco, CA 94112, USA\\
$^4$Department of Astronomy, Columbia University,
  New York, NY 10025, USA\\
  }

\date{Received ---- / Accepted ----}
\titlerunning{Stability of rotating circumstellar flows}

\abstract{ The hydrodynamic response of the inviscid small
shearing box model of a midplane section of a rotationally
supported astrophysical disk is examined.
An energy functional ${\cal E}$ is formulated
for the general nonlinear problem. It is found that the fate of disturbances is
related to the conservation of this quantity which, in turn, depends on the
boundary conditions utilized: ${\cal E}$ is conserved for
channel boundary conditions while it is not conserved
in general for shearing box conditions.
Linearized disturbances
subject to channel boundary conditions have normal-modes described
by Bessel Functions and are qualitatively governed by a quantity $\Sigma$
 {which is a measure  of the ratio between the azimuthal and vertical
wavelengths}.
Inertial oscillations ensue if $\Sigma >1$ - otherwise
disturbances must in general be treated as an initial value
problem.   We reflect upon these results
and offer a speculation.   }
  \maketitle

\begin{verse}``Extraordinary how mathematics help you to know yourself."\\
Samuel Beckett, \emph{Molloy}
\end{verse}

\section{Introduction}
The hydrodynamic response in small shearing box (SSB hereafter) models of
rotationally supported flows has been receiving renewed
attention over the past few years.  Because such fluid
configurations do not support supercritical linear instabilities
alternative forms of transition, possibly subcritical, have
been proposed, discussed and explored in numerous recent studies
(Ioannou \& Kakouris, 2001, Longaretti, 2002, Chagelishvilli \textit{et al.}, 2003,
Yecko, 2004, Mukhopadhyay \textit{et al.}
2004, Afshordi \textit{et al.} 2004, Umurhan \& Regev, 2004, Sternberg, 2005,
Lesur \& Longaretti, 2005, Balbus \& Hawley, 2006,
Rincon \textit{et al.} 2007, Ji \textit{et al.}, 2006, Williams, 2006\footnote{Note that this work
originally appeared in Williams (2000).}, Jim\'{e}nez \textit{et al.}, 2006,
Shen \textit{et al.}, 2006).
The identification of a route to sustained nonlinear activity for three dimensional
disturbances has been elusive for flows that are wall-bounded
(see recently Rincon \textit{et al.} 2007)
and for flows having open boundaries (Shen \textit{et al.}, 2006)
although weak levels of activity have been reported by
Lesur \& Longaretti (2005) near the Rayleigh line in flows
with open boundaries.\par
Far from answering whether or not such transitions
do exist in these flows, we instead
identify two clues which could be used to cast some light onto
this question in the future.
To this end we reconsider
the dynamics of inviscid anticylonic Rayleigh stable flows in the SSB model (\textit{i.e.},
rotating plane Couette flow, rpCf hereafter)
subject to
shearing boundary conditions (Goldreich \& Lynden-Bell, 1965,
Rogallo, 1981, Knobloch, 1984, Korycansky, 1992,  SBC hereafter) or channel boundary conditions (\textit{e.g.},
Yecko, 2004, Mukhopadhyay, et al. 2004, Lesur \& Longaretti, 2005, Rincon
et al. 2007, CBC hereafter).\par
The first result is a general nonlinear feature pertaining
to the energetics of such flows.
If one defines ${\cal E}$ as the total energy of the flow
\emph{in the SSB}, whose dynamical constituent includes
the total kinetic energy, (\textit{i.e.}, the
energy formed by considering the total
velocity - which is to say the velocity fluctuations plus
background shear flow), the fate of ${\cal E}$ depends
upon the boundary conditions
employed. ${\cal E}$ is conserved for (but not limited to)
(i) disturbances subject to CBC,
(ii) azimuthally and vertically periodic, radially localized disturbances in
radially unbounded domains.
On the other hand, disturbances subject to SBC
do not identically conserve ${\cal E}$ and
its temporal quality depends critically upon the dynamics within
the domain.  The nonconservation of ${\cal E}$ for flows
subject to SBC relates
to the feature that the kinetic energy
of a fluid parcel jumps when it crosses the (quasi) periodic radial boundaries.\par

For our second result, we find that there are qualitative differences in the
behavior of \emph{linearized disturbances} subject to CBC
 depending on whether
or not the quantity $\Sigma$ is less than or greater than 1.
This parameter is defined as
\beq
\Sigma = 4\frac{2(2-q)}{q^2}\frac{\ell_y^2}{\ell_z^2},
\eeq
where $\ell_z,\ell_y$ are
the disturbance length scales
in the vertical (like the disk normal) and
streamwise (like the disk azimuth) directions
of any given modal disturbance.
The global
flow for which the SSB is meant to represent
is characterized by a rotation rate $\Omega$
that depends on the radial coordinate $R$ according
to the relationship $\Omega \propto R^{-q}$.   {This
means that on the local scale there is a
Couette velocity profile given as $U = -q\Omega_0 x {\bf {\hat y}}$.}
Keplerian flow is when $q=3/2$.
Additionally and hereafter, we will consider
linear shear profiles that are anti-cyclonic, \textit{i.e.}, those
in which $q>0$. At infinite Reynolds numbers (Re $=\infty$) axisymmetric linear instability
(Taylor vortices) occurs for $q\ge 2$.
The significance
of $\Sigma$ was recognized in previous studies
(Knobloch, 1984, Korycansky, 1992, Sternberg, 2005,
as the parameter $\beta$ in Balbus \& Hawley, 2006)
of disturbances subject to SBC.   {In those studies
it was demonstrated that modes always decay.  If the mode
was initially leading then it will experience one maximum
whereas if the mode is trailing it never experiences
a maximum.  However, irrespective of this
the asymptotic decay of disturbances
depends on $\Sigma$:
for $\Sigma < 1$ the decay
is monotonic while for $\Sigma >1$ the decay
oscillatory.}

\par
In flows subject to CBC,
with given vertical and azimuthal wavenumbers,
if $\Sigma > 1$ the system
supports a countably infinite number of inertial normal modes described
by Bessel Functions
oscillating without growth or decay. Otherwise
 if $\Sigma < 1$ the system allows for
only one eigenmode - meaning that in general
the dynamical response cannot be characterized
by discrete normal modes - in which case,
such linearized disturbances must be treated as an initial value problem,
a situation well-known in studies of plane-Couette flow (Schmid \& Henningson, 2000,
pCf hereafter).
\par
  {We would like to note that the two main results
 we report upon here may be clues
 in explaining the results of Lesur \& Longaretti (2005) who show that
 there is a subcritical transition beyond the Rayleigh line
 in rpCf when SBC conditions are applied whereas no such
 transition was observed for such flows under CBC boundary
 conditions.  On a related note, as well, Lithwick (2007)
 demonstrated the existence of steady vortex structures in
 3D rpCf flow for Keplerian shear profiles when SBC are applied.
 We return to this in the discussion.}


\par

\section{Equations, Boundary Conditions and Two Integral Statements}
The equations we consider represent the dynamics taking place in a
``small-shearing box"  section located at a cylindrical
radius $R=R_0$
centered about the midplane of a
disk rotating about a central object with the local rotation
vector, $\Omega_0 {{\bf \hat z}}$.
With the external gravity set to zero and the density constant, these non-dimensionalized
equations (Goldreich \& Lynden-Bell, 1965,Umurhan \& Regev, 2004) are
\beqa
& & \nabla\cdot {\bf u} = 0, \label{ssb_incompressible}\\
& & (\partial_t - q\Omega_0x \partial_y) u + {\bf u}
\cdot\nabla u -2\Omega_0 v = -{\partial_x P}
\label{ssb_radial}\\
& & (\partial_t - q\Omega_0x \partial_y) v + {\bf u}
\cdot\nabla u + (2-q)\Omega_0u = -{\partial_y P},
\label{ssb_azimuthal}\\
& & (\partial_t - q\Omega_0x \partial_y) w + {\bf u}
\cdot\nabla w  = -{\partial_z P}.
\label{ssb_vertical}
\eeqa
The flow on these scales is
equivalently known as incompressible rpCf
(\textit{e.g.}, Nagata, 1986, Yecko, 2004). In this form these equations are the inviscid
limit of those considered by Longaretti (2002), Mukhopadyay et al. (2005) and Yecko (2004).
In the language of this paper,
$x$ corresponds to the radial (shearwise) coordinate,
$y$ corresponds to the azimuthal (streamwise) coordinate
and $z$ corresponds to the vertical (spanwise) coordinate -
with the corresponding
velocity disturbances ${\bf u} = \{u,v,w\}$.  These velocities represent
deviations over the steady Keplerian flow (as manifesting itself
in this rotating frame) given by $-q\Omega_0 x {\bf {\hat y}}$.
$\Omega_0$ is sometimes also referred to as the  Coriolis parameter,
and in these nondimensionalized units
$\Omega_0$ is $1$.
The local shear gradient
is defined as the exponent of the general
rotation law $\Omega(R) = \Omega_0 (R/R_0)^{-q}$.

\par
\emph{Boundary Conditions.} We first present and discuss the means by which
one administers {\em shearing boundary conditions} (SBC) in the SSB.
By this we mean to say
that the flow variables $\bf u$, and $p$ are (a) periodic in the
azimuthal and vertical directions on scales $L_y=1$ and $L_z=1$
respectively and (b) are simply periodic
(on scale $L_x=1$) in the
shearwise (radial) direction with respect to a coordinate frame
that moves with the local shear (e.g. Lynden-Bell \& Goldreich, 1965, Rogallo, 1981).
In this work we
refer to this coordinate frame as the {\em shearing coordinate} frame (SC
for short).  The
coordinate frame of an observer in the rotating frame,
will be referred to as the {\em
non-shearing coordinate frame} (NSC for short).
In this sense
the fundamental equations
(\ref{ssb_incompressible}-\ref{ssb_vertical}) have been expressed
in the NSC frame.\par
If $(X,Y,Z,T)$ represent the independent
variables in the SC frame then we say they are related to the NSC
variables by $X = x,  \ Y = y + q\Omega_0 x t,\ Z = z, \
T = t.$
If $f$ is
any dynamical quantity then in the SC frame the periodic boundary
conditions in the shearwise, azimuthal and vertical directions
respectively are
$ f(X,Y,Z) = f(X+L_x,Y,Z), \
f(X,Y,Z) = f(X,Y+L_y,Z),\
f(X,Y,Z) = f(X,Y,Z+L_z).$
Of course this means in general that $f(X,Y,Z) = f(X+L_x,Y+L_y,Z+L_z)$.
As expressed in the NSC frame, the periodicity in the
azimuthal and vertical directions appears like
$
f(x,y,z) = f(x,y+L_y,z), \
f(x,y,z) = f(x,y,z+L_z),
$
while the periodicity in the shearwise direction is
$
f(x,y,z) = f(x+L_x,y+q\Omega_0 t L_x,z).
$
Because of the periodicity in the azimuthal direction this
statement takes the form
$f(x,y,z) = f(x+L_x,y+\tilde y,z)$ where
$\tilde y(t) = q\Omega_0 t L_x ({\rm mod} \ L_y)$.  Note that this means
that only after {\em exactly} integer values of $q\Omega_0 t L_x/L_y$, (\textit{i.e.}, when $\tilde y = 0$)
are the SC and NSC frames coincident, meaning every
$ T_{{\rm rep}} = L_y/q\Omega_0 L_x,
$ time units.
Evidentally the SBC are time dependent
when viewed by an observer in the NSC frame.\par
When enforcing \emph{channel boundary conditions} (CBC)
on (\ref{ssb_incompressible}-\ref{ssb_vertical}) we require
that all quantities are periodic in the
vertical and azimuthal directions (in the NSC frame)
while requiring no normal-flow at the channel boundaries.  This
latter statement amounts to
$
u = 0$ at $x = 0,L_x$.
\par
\emph{A Pair of Integral Statements}.
The dynamical equations (\ref{ssb_incompressible}-\ref{ssb_vertical})
describe the evolution
of velocities which are disturbances about a steady
rotationally supported flow which, on the SSB scales, is
the linear shear.
Thus, the steady solution, ${\bf u} = 0$, is
interpreted as the undisturbed state.
We now consider the {\em total} velocity $\bf v$ defined as
$
{\bf v} \equiv -q\Omega_0 x {\bf{ \hat y}} + \bf u.
$
As such the governing equations of motion
(\ref{ssb_incompressible}-\ref{ssb_vertical})
are more concisely written in vector form,
\beqa
\partial_t{\bf v} + {\bf v}\cdot \nabla {\bf v} &=& -
\nabla p  - 2\Omega_0 {\bf{\hat z}}\times({\bf v} + q\Omega_0 x
{\bf{\hat y}}), \label{total_velocity_eqn} \\
\nabla\cdot{\bf v} &=& 0 \label{total_incompressibility}.
\eeqa
Note that these equations written in this way have appeared in other
works (\textit{e.g.}, Lesur \& Longaretti, 2005).
Taking the inner product of
(\ref{total_velocity_eqn}) with ${\bf v}$ and following
some manipulation
yields the following
\beq
{\partial_t}\varepsilon +
{\bf v}\cdot \nabla \left(\varepsilon + p\right) = 0,
\qquad
\varepsilon \equiv
\frac{{\bf v}^2}{2}
- q\Omega_0^2 x^2.
\label{F_eqn}
\eeq
With use of the incompressibility
condition (\ref{total_incompressibility})
we may integrate (\ref{F_eqn}) over the full spatial domain
to find,
\beq
\frac{d{\cal E}}{dt} = -\int_{{\cal S}}{(\varepsilon+p) {\bf v}\cdot {\bf \hat n}} dS,
\qquad
{\cal E} \equiv \int_{{\cal V}}\varepsilon dV,
\label{F_eval}
\eeq
in which
${\cal V}$ and ${\cal S}$ are the volume and surface-boundaries and
${\bf \hat n}$ is the unit normal of the bounding surface.
${\cal E}$ is comprised of (a)
the term ${\bf v}^2/{2}$
which is the kinetic energy and (b)
the potential-like term $-q\Omega_0^2 x^2$.
The global integral ${\cal E}$ can change due to
the influx of $\varepsilon$ across the boundaries , \textit{i.e.},
$\int_{{\cal S}}\varepsilon {\bf v}\cdot {{\bf \hat n}} dS$,
and through the external work done upon the system
denoted by
the boundary integral $\int_{{\cal S}}p {\bf v}\cdot {{\bf \hat n}} dS$.
The appropriate Bernoulli function for this system is
$\varepsilon + p$.
The functional ${\cal E}$ is analogous to the total energy contained
in the flow and its dynamical response (see below) is characterized by the
total kinetic
energy variations in the box.
\par
By contrast we follow the same procedure as above but consider
an energy functional comprised only of the disturbance velocities.
Beginning with
(\ref{ssb_radial}-\ref{ssb_vertical})
and taking its inner product with the disturbance velocity
${\bf u}$, followed by the use of (\ref{ssb_incompressible}) and either the
SBC or
CBC boundary conditions, one can integrate over the domain
to find
\beq
\frac{dE}{dt} \equiv \frac{d}{dt}\int_{{\cal V}}\xi dV = q\Omega_0\int_{{\cal V}} uv dV; \qquad
\xi \equiv \sfrac{1}{2}{\bf u}^2.
\label{ReynoldsOrr}
\eeq
The disturbance energy per unit volume is $\xi$.  The total domain integrated
\emph{disturbance energy} is $E\equiv\int_{{\cal V}}\xi dV$.
The above is known in more general terms as the Reynolds-Orr equation
(e.g. Schmid \& Henningson, 2000)
and relates
the change of the total disturbance energy to the domain integrated correlation between
$u$ and $v$.
In the absence of shear (\textit{i.e.}, $q=0$), the disturbance energy
is conserved.
\par
Inspection of the Reynolds-Orr
relationship says that
so long as
there is some correlation between the radial and azimuthal
velocities the disturbance energy must fluctuate in time.
However from the definitions of
${\cal E}$ and $E$ we have
\beq
{\cal E} = E + E_{{\rm shear}}; \
E_{{\rm shear}} \equiv
\int_{{\cal V}}\left[{-q\Omega_0 x v + \sfrac{1}{2}q^2\Omega_0^2 x^2}\right]dV,
\label{def_E_shear}
\eeq
we see that disturbance energy $E$
is related to ${\cal E}$ (the global energy) in an intimate way via
$E_{{\rm shear}}$, which we consider to be the disturbance energy
affected by the shear in a dynamical sort of way.
\par
To appreciate the significance of these relationships
we first apply CBC to
(\ref{F_eval}) and find that
\beq \frac{d{\cal E}}{dt} = 0.
\label{calE_conservation} \eeq
This means that the
fluctuations in $E$ come at the expense of $E_{{\rm shear}}$ in an
equal but opposite way throughout the dynamical response of the
flow; \textit{i.e.}, the stresses comprising the RHS of
(\ref{ReynoldsOrr}) is a measure of the time rate of change
between these two forms of measured energy.
\par
We note that if instead
of the CBC we require that the flow
be periodic in $y$ and $z$ (again) but that all flow quantities decay
sufficiently fast as $x\rightarrow\pm\infty$, then the boundary term
in (\ref{F_eval}) also vanishes and (\ref{calE_conservation}) again follows.  The
same is true if the vertical domain is unbounded and disturbances are vertically
localized as well, i.e.
(\ref{calE_conservation}) also follows if
disturbances similarly decay
as $z\rightarrow\pm\infty$.
\par
Matters are less clear when SBC are employed because
${\cal E}$ is no longer
conserved, evolving according to
( {with boundaries of the periodic box set at $x=\pm \sfrac{1}{2}$}),
\beq \frac{d{\cal E}}{dt} =
q\Omega_0 \int_{{\cal S}_{1}}uv dy dz, \label{phi_evolution_sc}
\eeq in which ${\cal S}_{1/2}$  {is the boundary surface at $x = 1/2$}.
In general the right-hand side of
(\ref{phi_evolution_sc}) is not expected to be zero and it means that
${\cal E}$ can vary with respect to time when SBC are employed.
The physical reason for this is clear:
 {\emph{as a fluid parcels exits
from one radial boundary with a given azimuthal velocity
it reenters through the other boundary with
an azimuthal velocity boosted by the difference in the background
velocities between the two radial boundaries.}
Another observation about this is that
disturbances are forced \emph{a-priori} to have a correlation
scale set by the length of the box $L$ (Regev \& Umurhan, 2008).}
We note that ${\cal E}$ will be conserved for SBC if the disturbances
are sufficiently localized in the interior of the computational
domain so that there is no power (i.e. fluid motion)
near the bounding surface ${\cal S}_{1/2}$ (Nusser, 2007).

\section{Linear Analysis with CBC}
In considering the response of the flow in the channel model we
shall assume initially the general form
$
 f(x,y,z,t)=f_{_{\alpha \beta}}(x,t)\exp ({i\alpha y+i\beta z})+c.c.$
Linearization of (\ref{ssb_incompressible}-\ref{ssb_vertical}) followed
by introduction of this spatial Fourier form and rearranging
results in the single equation for the radial velocity
\beq (\partial_t -i q \Omega_0 x\alpha)^2
(\partial^2_x - \beta^2 - \alpha^2) u_{_{\alpha\beta}} =
\omega_e^2 \beta^2 u_{_{\alpha\beta}}, \label{uber_master_ux_eqn}
\eeq
 {in which the epicyclic frequency is defined by
$\omega_e^2 \equiv 2(2-q)\Omega_0^2$.}
A full breadth of
normal-mode solutions of the form $\exp(i\omega t)$ exist only for
certain combinations of $\beta$ and $\alpha$. To see what the
restriction is in the following we perform a normal analysis and
determine precisely under what conditions this circumstance
occurs.  Thus, assuming the normal mode form $
u_{_{\alpha\beta}} = \hat u_{_{\alpha\beta}} \exp{i\omega t},
$
into (\ref{uber_master_ux_eqn}) results in
 \beq (\omega - q \Omega_0 x\alpha)^2
(\partial^2_x - \beta^2 - \alpha^2)\hat u_{_{\alpha\beta}} +
\omega_e^2 \beta^2 \hat u_{_{\alpha\beta}} =0. \label{master_ux_eqn}
\eeq
 {The solution
to (\ref{master_ux_eqn}) subject to CBC }is given by
\beqa  \hat u_{_{\alpha\beta}} &=& A(\omega
- q \Omega_0 x  \alpha)^{1/2} \Bigl\{ J_{\nu}\bigl[i\lambda(\omega -
q\Omega_0 x \alpha)\bigr]
Y_{\nu}\bigl[i\lambda\omega\bigr] -  \nonumber  \\
& & Y_{\nu}\bigl[i\lambda(\omega - q\Omega_0 x \alpha)\bigr]
J_{\nu}\bigl[i\lambda\omega\bigr]\Bigr\}, \label{3d_channel_solution}
\eeqa where
\beq \lambda^2 = \frac{\beta^2 +
\alpha^2}{q^2\Omega_0^2\alpha^2},\qquad \nu^2 = \frac{1 -
\Sigma}{4}, \qquad
\Sigma = \frac{4\omega_e^2\beta^2}{q^2\Omega_0^2\alpha^2}.
\eeq
where $J_\nu$
and $Y_\nu$ are Bessel Functions of order $\nu$.
The quantization condition on $\omega$ is
\beqa (\omega-\omega_{\infty})^{1/2} \Bigl\{ J_{\nu}\bigl[i\lambda(\omega-\omega_{\infty})\bigr]
Y_{\nu}\bigl[i\lambda\omega\bigr] \ \ \ \ \ \ \ \ \ \  & & \nonumber \\
- Y_{\nu}\bigl[i\lambda(\omega-\omega_{\infty})\bigr]
J_{\nu}\bigl[i\lambda\omega\bigr] \Bigr\}
 &=& 0. \eeqa
The governing
equation and its solution has a regular singular point for $\omega
= \omega_\infty \equiv q\Omega_0\alpha$ at the point $x=1$.  As it
turns out, the solution automatically satisfies its requisite
boundary condition at $x=1$ when $\omega = \omega_\infty$.
These modes are the fundamental structures that
constitute the so-called \emph{wallmodes} that
give rise to the stratorotational instability
(Yavneh \textit{et al.}, 2001, Dubrulle \textit{et al.}, 2005 Umurhan, 2006).
\par
We observe that if $\nu$, as appearing
, is a real quantity then there is no way for the quantization
condition to be met by real values of $\omega$ (other than the
single value $\omega_\infty$) according to the known behavior of
the Bessel functions (Abramowitz \& Stegun, 1972).  This means
that if $\Sigma < 1$, then {\em there is only one normal mode
solution for this problem}.
For $\Sigma > 1$  there exists a countably infinite
set of normal-mode solutions (overtones) as per the theory of
Bessel functions (see again Abramowitz \& Stegun, 1972).
\par
Nevertheless, inferring
the dynamical content of the disturbances, including the shape of
the eigenfunctions and the general formulae for the dispersion
relationship, is not easy. One can more transparently study the
behavior of these features, including their character change, if an asymptotic solution for
$u_{_{\alpha\beta}}$ is developed instead for
the small limiting values of $\beta$ and $\alpha$.
We proceed
by assuming $\epsilon \ll 1$
and,
$
\beta = \epsilon \beta_1, \  \alpha = \epsilon\alpha_1,
$  followed by the expansion procedure
$
\omega = \epsilon\omega_1 + \epsilon^3\omega_3 + \cdots$
and
$
\hat u_{_{\alpha \beta}} = u_0 + \epsilon^2 u_2 + \cdots,
$
to the solution of (\ref{master_ux_eqn}).
We find that the lowest order solutions depend critically on the
parameter $\Sigma_1$ defined by $
\Sigma_1 = {4\omega_e^2\beta_1^2}/{q^2\Omega_0^2\alpha_1^2},$
which is really a reexpression of $\Sigma$.
The leading order allowed solutions
is $u_0$
\beq u_0 = A_0\left(1 - \frac{q\Omega_0 x
\alpha_1}{\omega_1}\right)^{1/2}\sin \left [\delta \ln\left (1 -
\frac{q\Omega_0 x \alpha_1}{\omega_1} \right) \right ],
\label{asymptotic_solution}
\eeq
with $
2\delta = \sqrt{1 - \Sigma_1}$ and
where $A_0$ is an arbitrary
amplitude.
The asymptotic form
predicts the absence of normal-mode
solutions if $\Sigma_1 < 1$ except for one with $\omega_1 = q\Omega_0\alpha_1$
(the wallmode).
The quantization condition on
$\omega_1$, resulting from imposing the boundary
condition at $x=1$ results in
$ \omega_1 = {q\Omega_0
\alpha_1}/[1-\exp({n\pi}/{|\delta|})], \ \ n = \pm 1, \pm 2,
\cdots$,
where $n$ labels the particular overtone in question.
There are an infinite number of radial overtones for any
allowable pair of $\beta_1$'s and $\alpha_1$'s.\par
\par
In a more general sense:
when $\Sigma<1$, aside from the mode $\omega=\omega_\infty$,
the response of the flow must be considered as an initial value problem
since the system supports no other discrete normal modes.  When $\Sigma>1$
then there are two branches of normal modes (\textit{i.e.}, overtones) clustering about
$\omega = 0$ and $\omega=\omega_\infty$ - in general the frequencies
around which the overtones cluster
correspond
to wavespeeds (\textit{i.e.}, $c=\omega/\alpha$) that equal the flow speeds at the
two channel boundaries.
\par
Modes with $n>0$ are the symmetric brethren of those
with $n<0$.  The reason is that (\ref{master_ux_eqn})
remains unchanged
after applying the spatial reflection operation about the point $x=1/2$
together with the frequency reflection/shift $\omega \rightarrow -\omega + q\Omega_0 \alpha$.
A given eigensolution $\hat u^{(+)}_{\alpha\beta}(x)$ with eigenfrequency $\omega_{+}$
generates another eigensolution $\hat u^{(-)}_{\alpha\beta}$
with eigenfrequency $\omega_{-} = -\omega_{+} + q\Omega_0\alpha$
such that $\hat u^{(-)}_{\alpha\beta}$ is the mirror reflection
of $\hat u^{(+)}_{\alpha\beta}$ about $x=1/2$.
In general we see that the approximate theory - including the
predicted eigenfrequencies -  are good even for order 1 values of
$\alpha$ and $\beta$, but the theory begins to breakdown when they
exceed order 1 values (Figure 1).

\begin{figure}
\begin{center}
\leavevmode \epsfysize=5.45cm
\epsfbox{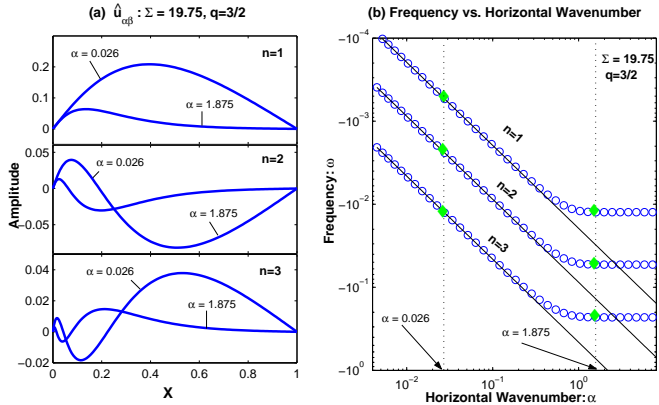}
\end{center}
\caption{{\small Results from normal mode theory in which
$\Sigma =19.75$ and $q=3/2$ and $\Omega_0 = 1$.  (a) The exact eigenfunctions for
$\hat u_{\alpha\beta}$ depicted for the
first three overtones each at two values of $\alpha$.
The eigensolutions developed via the asymptotic
theory lie exactly on the eigenfunctions for $\alpha = 0.02$ where
$A_0 = 1/2\delta$.
(b) The frequencies $\omega$ as a function of $\alpha$ for the
first three overtones are shown (open circles).  The predicted frequencies
using the asymptotic theory are shown with solid lines.
The values corresponding to the eigenfunctions depicted in the left
panel are designated with diamonds.  Note that only the positive
overtones (\textit{i.e.}, $n>0$) are shown here - the negative overtones
are reflection symmetric about $x=1/2$ as discussed in the
text.
}} \label{normal_modes}
\end{figure}

\section{Reflections and Speculation}
\emph{1. A Speculation.}
With respect to the question of dynamics of wall-bounded rpCf,
these results are quite suggestive.  Rincon et al. (2007) showed
that Rayleigh-stable anticyclonic rpCf appears not to
show a transition into a turbulent state in the manner
in which transition is observed
in pCf flow (\textit{e.g.}, Waleffe, 2003).  The strategy in those
problems is to identify \emph{steady} nonlinear solutions of the flow
(usually by some continuation method from known solutions
in otherwise unstable parameter regimes, \textit{e.g.},  Nagata, 1990, or forcing
methods, Waleffe, 2003) and study
their stability properties.   If there are a collection of steady nonlinear
structures, some of which
unstable, then the transition into a turbulent
state is thought to be approached.  These states get
easily triggered (bypass mechanism) because
small scale disturbances in pCf get amplified
due to the lift-up effect -  {a mechanism by which axisymmetric
disturbances result in algebraic instability (Ellingsen \& Palm, 1975).}
Rincon et al. (2007) however show that there are two problems
with this approach when employed to identify
structures of rpCf near the Rayleigh-stable line (\textit{e.g.}, near
 $q=2$ for Re $\rightarrow \infty$):
that steady structures could not be identified and
 {that the anti lift-up effect, which is
the lift-up effect analog on the Rayleigh line, induces radial/vertical velocity fields
that do not undergo the sorts of secondary instabilities
that are characteristic of
the induced azimuthal velocity
fields generated by the lift-up effect in pCf flow.}\par

Since the linear analysis of the inviscid Rayleigh-stable rpCf
shows that for a given $\alpha$ and $\beta$ there are conditions
where there exists a countable number of oscillating modes,
\emph{we openly speculate here} that one might
consider identifying oscillating nonlinear structures instead
of steady ones.    Perhaps
one strategy would be to (i) identify oscillating
nonlinear states in the infinite Reynolds Number case (Re $=\infty$)
and then (ii) ratchet down Re and follow/study the structures
and their activity until either (iii) one identifies a Re of transition,
if a unique one exists free of hysteretic effects or
determine if,
as Rincon \textit{et al.} (2007) suggest,
 rpCf exhibits the character of a chaotic
saddle.  However we note that identifying
oscillating (travelling wave) nonlinear solutions
is difficult in plane Poiseuille flow (pPf) as Toh \& Itano (2003)
have recently shown.  However,
\emph{it might} be easier to identify such structures in
rpCf because of the presence there of linear oscillations
whereas such oscillations are absent in linearized pPf.
\par
\emph{2. Some Reflections.}
An individual wavemode subject to SBC, i.e.
$
\sim \exp{(ikX + i\alpha Y + i\beta Z)},
$
where $k,\alpha$ and $\beta$ are wavenumbers as before,
was shown by Balbus \& Hawley (2007) to be an exact
solution of the nonlinear equations of motion
(i.e. \ref{ssb_incompressible}-\ref{ssb_vertical}).   {This
is because this SBC waveform ansatz, together with the
incompressibility condition, results in the
nonlinear terms exactly cancelling to zero for
each velocity component.}
Given (\ref{ReynoldsOrr}-\ref{def_E_shear})
and
(\ref{phi_evolution_sc})
it follows
that for these solutions the following holds,
\[
\frac{d{\cal E}}{dt} = \frac{dE}{dt},
\]
where $E$ is the disturbance kinetic energy of the
wavemode
(see \ref{ReynoldsOrr}).  From
(\ref{def_E_shear}) it follows that these
disturbances conserve the shear energy $E_{{\rm shear}}$.
 {A study of the form of $E_{{\rm shear}}$ shows that
this is a natural outcome due to the symmetry of
the modes with $\alpha,\beta \ne 0$.  However, for modes
in which $\alpha = \beta = 0$, an inspection of
the governing equations of motion in the SC shows
that it necessarily follows that the corresponding
azimuthal velocity mode, i.e. which is uniform in the
vertical and azimuthal directions, is zero.
Thus, by its very construction, these exact nonlinear modes
as a result of SBC cannot support nonlinear
modifications to the shear. Consequently, such alterations can occur only
via three-wave interactions (Balbus \& Hawley, 2007).
We note that no such disturbances are supported in
CBC flows - there always exists some amount of self-interaction
for most modes of the system.
}
\par
 {
We note that for a general collection
of perturbations utilizing SBC,
(\ref{phi_evolution_sc}) says that
$\dot {\cal E} \ne 0$.  The fluctuations in the
total energy relate directly to the lateral exchange of
kinetic energy from radially
neighboring copies of the shearing box,
where azimuthal velocities enter or exit boosted
by the difference in the background shear velocities
between the two boundaries -
the quality of this
exchange is a result of the type of correlations imposed upon
the disturbances.  For SBC it means that
these correlations are set a priori by the periodic length
scale of disturbances which is set to be on the scale of the box
$L$ (e.g. see Regev \& Umurhan, 2008).  By contrast, the
imposition of no-normal flow boundary conditions removes
this source of energy into and out of the domain
for problems invoking CBC, thus $\dot {\cal E} = 0$ in
that case.
}
\par
 {We feel that these observations about the differences between
CBC and SBC, together with recent results, raises more
questions that need to be resolved.
For instance, linear disturbances
of the CBC type are understood not to result in decay
because of reflections off of the channel walls.
SBC conditions, on the other hand, support modes which
can result in fluctuations of the total energy
depending on the correlations that exist across
the radial boundaries.  It is interesting, therefore,
that Lesur \& Longaretti (2005) demonstrate the persistence
of activity for simulations of the SSB (slightly
below the Rayleigh line) using SBC while they report
the presence of no such activity for CBC.
On the other hand, there is in Shen et al. (2006) a
recent demonstration of the decaying quality of disturbances
in the SBC for flows run with $q=3/2$.
\footnote{
Although
compressibility effects are included.}
Equally intriguing is that
Lithwick (2007) shows that a single vortex may persist
for a long time in a three-dimensional simulation of the SSB
for $q=3/2$ using SBC which involves the
complex interplay of three-wave mode interactions.
It is interesting to us that while CBC conserves
its total energy ${\cal E}$ no activity has yet been observed
under Keplerian-like
conditions while for SBC, which does not
conserve its ${\cal E}$,
sustained activity occurs in some cases
while decay characterizes others.}
\par
 {Given the preceeding discussion it seems there are
important questions to be asked:
\begin{enumerate}
\item What is the role
of the non-self interacting
wave modes resulting from the use
of the SBC in the maintenance and/or destruction
of activity in SSB simulations?
\item What influence does the prescribed velocity correlations,
which result in the total energy not being conserved
(cf. (\ref{phi_evolution_sc}))
have on the overall evolution of general SSB simulations using
SBC?
\item What influence does the non-self-interacting nature
of these waves in the SBC have on
the velocities on the
radial boundaries of the system which, in turn, affects
the amount of energy entering/leaving the system.?
\item How are we to understand these effects in relation
to dynamics supported in the nonlinear
response of flows utilizing CBC (i.e. wall-bounded rpCf)?
\end{enumerate}
These observations and questions underscore the need for further analysis
(also see Dubrulle, \textit{et al.}, 2005).}
\par
\emph{Acknowledgements}.  We thank the
anonymous referee who gave us critical comments
that helped us focus the presentation of this work
and pointed us to earlier references.
This work was partially
supported by BSF grant number 0603414082.


\end{document}